\documentclass[11pt,onecolumn,floatfix,altaffilletter,superscriptaddress,tightenlines,showpacs,showkeys,preprintnumbers,nofootinbib]{revtex4-1}
\pdfoutput=1
\usepackage{hyperref}
\usepackage{amsmath,amssymb}
\usepackage{epsfig} 
\usepackage{graphicx}        
\usepackage{url}
\usepackage{color}
\usepackage{multirow}
\usepackage{placeins}
\usepackage[dvipsnames]{xcolor}
\usepackage{braket}
\usepackage{float}
\usepackage{slashed}
\usepackage{fdsymbol}
\usepackage{MnSymbol,bbding,pifont,array}
\hypersetup{colorlinks=true,linkcolor=blue,citecolor=teal,filecolor=magenta, urlcolor=purple}

\allowdisplaybreaks
\newcolumntype{B}[1]{>{\centering\arraybackslash}m{#1}}
\bibpunct{[}{]}{,}{n}{}{,}

\newcommand{\INFN}{INFN - Sezione di Napoli, Complesso Universitario Monte S. Angelo, I-80126 Napoli, Italy}
\newcommand{\UNINA}{Dipartimento di Fisica ``Ettore Pancini", Università degli studi di Napoli ``Federico II'', Complesso Universitario Monte S. Angelo, I-80126 Napoli, Italy}
\newcommand{\SSM}{Scuola Superiore Meridionale, Via Mezzocannone 4, 80138 Napoli, Italy}
\newcommand{\NNU}{Institute of Theoretical Physics and Institute of Physics Frontiers and Interdisciplinary Sciences, Nanjing Normal University, Wenyuan Road, Nanjing, Jiangsu, 210023, China}
\newcommand{\MEKL}{Nanjing Key Laboratory of Particle Physics and Astrophysics, Nanjing, 210023, China}

\begin{document}
\title{Impact of memory-burdened black holes on primordial gravitational waves in light of Pulsar Timing Array}

\author{Peter Athron}
  \email{peter.athron@njnu.edu.cn}
  \affiliation{\NNU}
  \affiliation{\MEKL}
\author{Marco Chianese}
  \email{marco.chianese@unina.it}
  \affiliation{\UNINA}
  \affiliation{\INFN}
\author{Satyabrata Datta}
  \email{amisatyabrata703@gmail.com}
  \affiliation{\NNU}
  \affiliation{\MEKL}
\author{Rome Samanta}
  \email{samanta@na.infn.it}
  \affiliation{\SSM}
  \affiliation{\INFN}
\author{Ninetta Saviano}
  \email{nsaviano@na.infn.it}
  \affiliation{\INFN}
  \affiliation{\SSM}
  
\begin{abstract}
Blue-tilted Gravitational Waves (BGWs) have been proposed as a potential candidate for the cosmic gravitational waves detected by Pulsar Timing Arrays (PTA). In the standard cosmological framework, BGWs are constrained in their frequency range by the Big Bang Nucleosynthesis (BBN) limit on GW amplitude, which precludes their detection at interferometer scales. However, introducing a phase of early matter domination dilutes BGWs at higher frequencies, ensuring compatibility with both the BBN and LIGO constraints on stochastic GWs. This mechanism allows BGWs to align with PTA data while producing a distinct and testable GW signal across a broad frequency spectrum. Ultralight Primordial Black Holes (PBHs) could provide the required early matter-dominated phase to support this process. Interpreted through the lens of BGWs, the PTA results offer a way to constrain the parameter space of a new scenario involving modified Hawking radiation, known as the ``memory burden" effect, associated with ultralight PBHs. This interpretation can be further probed by high-frequency GW detectors.
Specifically, we demonstrate that PBHs as light as $10^{2-3}~{\rm g}$ can leave detectable imprints on BGWs at higher frequencies while remaining consistent with PTA observations.
\end{abstract}
\maketitle
\tableofcontents

\section{Introduction} 

The concept of Primordial Black Holes (PBHs)~\cite{Hawking:1971ei, Carr:1974nx, Escriva:2022duf}, explored for over half a century, has been revitalized after the LIGO and VIRGO collaboration's groundbreaking detection of Gravitational Waves (GWs) from black hole mergers. PBHs, spanning masses from $M_{\rm PBH}\sim 0.1$ g to hundreds of solar masses, exhibit fascinating physical properties. In particular, long-lived PBHs within the asteroid mass range $10^{17}\lesssim M_{\rm PBH}/\text{g}\lesssim 10^{22}$ potentially comprise a significant portion of dark matter~\cite{Carr:2016drx, Green:2020jor, Carr:2021bzv}, while those with intermediate masses $10^{9}\lesssim M_{\rm PBH}/\text{g}\lesssim 10^{17}$ are subject to several constraints~\cite{Carr:2009jm, Carr:2020gox}. Ultralight PBHs with masses below $M_{\rm PBH}\lesssim 10^9\,\text{g}$ could have instead dominated the energy density of the Universe even though they undergo complete evaporation prior to Big Bang Nucleosynthesis (BBN) via Hawking Radiation (HR). This results in a plethora of potential effects and observable signatures such as generating beyond the Standard Model states including dark matter particles~\cite{Bernal:2020kse, Gondolo:2020uqv, Cheek:2021odj,  Samanta:2021mdm,  Bernal:2021bbv, Sandick:2021gew, Gehrman:2023qjn, Bertuzzo:2024fns}, producing the baryon asymmetry of the Universe~\cite{Fujita:2014hha, Hamada:2016jnq, Morrison:2018xla, Chen:2019etb, Perez-Gonzalez:2020vnz, Datta:2020bht, Hooper:2020otu, JyotiDas:2021shi, DeLuca:2021oer, Borah:2022iym, Calabrese:2023key, Calabrese:2023bxz, Schmitz:2023pfy, Barman:2024slw, Gunn:2024xaq}, sourcing and imprinting the Stochastic Gravitational Waves Background (SGWB)~\cite{Borah:2022vsu, Borah:2023iqo, Papanikolaou:2020qtd, Domenech:2020ssp, Papanikolaou:2022chm, Ireland:2023avg, Domenech:2024wao, Datta:2024bqp}.

The theoretical and phenomenological framework outlined above relies on a semiclassical treatment of black-hole evaporation, which has been recently challenged~\cite{Dvali:2018xpy, Dvali:2020wft, Dvali:2024hsb}. Hawking's semiclassical calculations for 4-dimensional black holes are expected to be invalid when the black hole has lost approximately half of its initial mass. This is primarily due to the neglection of a back-reaction, wherein the quantum information retained in the system's memory (or an excess of entropy) exerts a reciprocal effect on the black hole. This effect, also known as the ``memory burden", leads to a significant extension of the black hole's lifetime due to the accumulation of this quantum information or memory within the black hole. Such a pivotal deviation of the PBH evaporation from the semiclassical approximation yields profound phenomenological consequences which have been investigated in the context of dark matter~\cite{Dvali:2021byy, Alexandre:2024nuo, Thoss:2024hsr, Haque:2024eyh, Chianese:2024rsn, Zantedeschi:2024ram, Barker:2024mpz, Borah:2024bcr} and primordial GWs sourced by PBHs~\cite{Balaji:2024hpu, Barman:2024iht, Bhaumik:2024qzd, Barman:2024ufm, Kohri:2024qpd, Jiang:2024aju, Barker:2024mpz, Loc:2024qbz}.

In this article, we adopt a phenomenological approach to investigate the imprints of ultralight, memory-burdened PBHs on the SGWB generated by cosmic inflation~\cite{Guth:1980zm, Linde:1981mu}, parametrized as $\Omega_{\rm GW}\propto f^{n_T}$, with $n_T>0$. We shall refer to these GWs as the  Blue-tilted Gravitational Waves (BGWs). The primary motivation for our approach is twofold. First, any post-inflationary Early Matter-Dominated (EMD) phase suppresses power-law GWs through entropy production, which broadens the overall spectrum across a wide frequency range while remaining consistent with both the BBN and LIGO bounds on GW amplitudes. Moreover, the beginning and the end of the EMD phase leave distinct imprints on the GW spectrum. Ultralight PBHs can drive this necessary EMD phase, and the parameters governing these PBHs—such as those related to the memory burden effect—can be reconstructed from characteristic GW signals spanning multiple frequency bands. Second, the broad frequency range of the GW spectrum enables observational constraints in one frequency band to provide robust forecasts for other parts of the spectrum. The BGWs are considered to be one of the most promising cosmological sources of the recently detected nanohertz ($f\sim 10^{-9}$ Hz) SGWB by the Pulsar Timing Array (PTA) collaborations~\cite{NANOGrav:2023gor, NANOGrav:2023hvm, EPTA:2023fyk, Reardon:2023gzh, Xu:2023wog, EPTA:2023xxk, Vagnozzi:2020gtf, Bhattacharya:2020lhc, Kuroyanagi:2020sfw, Benetti:2021uea, Vagnozzi:2023lwo}.   Thus, the presence of ultralight PBHs in the early universe could generate a well-defined BGW signal spanning a broad frequency range.  Building on these two key motivations, our study pinpoints memory burden effects in PBHs as a potential mechanism for explaining the PTA data with BGWs, while simultaneously circumventing constraints from BBN and LIGO that pose significant challenges for conventional PBH scenarios with extremely light PBH masses. Additionally, it predicts distinct, high-frequency GW signals carrying imprints of the memory burden effect in PBHs, offering a compelling avenue for future observational tests.

A secondary motivation for our approach is that these memory-burdened PBHs could act as additional sources of gravitational waves (GWs). For instance, they might emit gravitons, contributing to an ultrahigh-frequency GW background~\cite{Anantua:2008am, Dolgov:2011cq, Dong:2015yjs}. This characteristic sets them apart from other forms of early matter domination (EMD), such as those driven by scalar fields, which could potentially yield similar spectral features on BGWs at lower frequencies~\cite{DEramo:2019tit,Datta:2022tab, Datta:2023vbs,Chianese:2024nyw}.

The paper is organized as follows. In Sec.~\ref{sec:evap}, we discuss the memory burden associated with PBHs and its impact on their lifetime. We then discuss in Sec.~\ref{sec:bgw} how such an impact on PBH lifetime is associated with the imprinting of BGWs from inflation. In Sec.~\ref{sec:results} we present our numerical results that are consistent with all constraints while focusing on a fit to the PTA data. Finally, in Sec.~\ref{sec:concl} we draw our conclusions.


\section{Impact of memory burden on PBH evaporation dynamics \label{sec:evap}}

In this section, we introduce the ingredients required to understand the impact of memory burden on PBH evaporation and its effects on the dynamics of a PBH-dominated universe. According to Ref.s~\cite{Dvali:2018xpy, Dvali:2018ytn, Dvali:2024hsb}, the back-reaction leading to the memory burden suppresses the evaporation rate by inverse powers of the black hole entropy $S^{k}$, where the exponent $k$ fixes the efficiency of back-reaction after the breakdown of the semiclassical approximation. We consider $k$ as a free parameter with $k\geq 0$. Due to memory burden, the semiclassical regime is valid until the threshold mass $M_{\rm mb}=q M_{\rm PBH}$ is reached, where $0<q<1$ and $M_{\rm PBH}$ is the initial PBH mass at formation. The PBH mass loss rate due to Hawking radiation is given by
 \begin{equation}\label{eq1}
     \frac{{\rm d}M_{\rm PBH}}{{\rm d}t}= -\frac{\mathcal{A} M_{\rm Pl}^4}{ M_{\rm PBH}^2}\times \left\{\begin{array}{l r}
     1 &\quad \text{for}\: M_{\rm PBH}\geq M_{\rm mb} \\
     S^{-k}(M_{\rm PBH}) &\quad \text{for}\: M_{\rm PBH}<M_{\rm mb}
     \end{array}
     \right. \,,
 \end{equation}
where $\mathcal{A}=\mathcal{G}\,g_{*}(T_{\rm PBH})/({30720\pi})$ with $\mathcal{G}\simeq 3.8$ being the greybody factor, $g_{*}(T_{\rm PBH})\simeq 108$ is the number of relativistic degrees of freedom below the Hawking temperature $T_{\rm PBH}$, and $S(M_{\rm PBH})=4\pi M_{\rm PBH}^2/M_{\rm Pl}^2$ with $M_{\rm Pl}=1.22\times 10^{19}~{\rm GeV}$ is the black-hole entropy. When memory burden sets in at $t_{\rm mb} = (1-q^3)t_{\rm ev}^{\rm sc}$, where $t_{\rm ev}^{\rm sc}=M_{\rm PBH}^3/ (3\mathcal{A} \,M_{\rm Pl}^4)$ is the semiclassical evaporation time, the evaporation dynamics starts to be modified allowing for a prolonged PBH lifetime before complete evaporation\footnote{A precise determination of the final fate of memory-burdened PBHs requires a fully quantum gravitational treatment, which remains beyond the reach of current theoretical methods. Nevertheless, given the possibility for classical instability or continued evaporation at a suppressed rate \cite{Dvali:2020wft}, assuming full evaporation provides a well-motivated working hypothesis for the purposes of our analysis.}. The total evaporation time $t_{\rm ev}$ can then be split up into,
\begin{equation} \label{eq2}
    t_{\rm ev}= t_{\rm mb} + \tilde{t}_{\rm ev}\,,
\end{equation}
where $\tilde{t}_{\rm ev}$ is the time from when memory burden effects set in until evaporation, and is equal to
\begin{equation}
\tilde{t}_{\rm ev} = \frac{3}{3+2k} q^{3+2k}S^{k}(M_{\rm PBH}) t_{\rm ev}^{\rm sc}\,.
\end{equation}
In the numerical computation, we shall consider that complete PBH evaporation occurs before BBN, i.e. $t_{\rm ev}\leq t_{\rm BBN} \sim 10^{25}\: \text{GeV}^{-1}$ \cite{bbn1,bbn2,bbn3}.

The dynamical evolution of energy densities of PBHs ($\rho_{\rm PBH}$) and radiation ($\rho_{\rm rad}$) is dictated by the following Friedmann equations
 \cite{Datta:2020bht,Masina:2020xhk}:
\begin{equation}
\frac{{\rm d}\rho_{\rm rad}}{{\rm d}z}+\frac{4}{z}\rho_{\rm rad}=0\,,\label{den1}
\end{equation}
\begin{equation}
\frac{{\rm d}\rho_{\rm PBH}}{{\rm d}z}+\frac{3}{z}\frac{H}{\tilde{H}}\rho_{\rm PBH}-\frac{\dot{M}_{\rm PBH}}{M_{\rm PBH}}\frac{1}{z\tilde{H}}\rho_{\rm PBH}=0\,,\label{den2}
\end{equation}
where $H$ is the Hubble parameter, $z=T_{\rm form}/T$ with $T_{\rm form}$ being the PBH formation temperature, and  $\tilde{H}=\left(H+\mathcal{K}\right)$ with $\mathcal{K}=\frac14\frac{\dot{M}_{\rm PBH}}{M_{\rm PBH}}\frac{\rho_{\rm PBH}}{\rho_{\rm rad}}$. The scale factor $a$ evolves according to 
\begin{equation}
\frac{{\rm d}a}{{\rm d}z}=\left(1-\frac{\mathcal{K}}{\tilde{H}}\right)\frac{a}{z}\,. \label{temvar}
\end{equation}
In the derivation of Eq.s~\eqref{den1}-\eqref{temvar}, we have assumed that the number of entropy ($g_{*s}$) and the energy ($g_{*\rho}$) degrees of freedom are equal and constant. We shall always work with the fractional energy density of PBHs, denoted as $\beta= {\rho_{\rm PBH}(T_{\rm form})}/{\rho_{\rm rad}(T_{\rm form})}$, such that they dominate the total energy budget in the early universe at a temperature $T_{\rm dom}=\beta T_{\rm form}$ (start of PBH-induced EMD) until they evaporate at $T\sim T_{\rm ev}$. For a given $\beta$, the above equations can be solved numerically to precisely determine the temperatures of PBH  evaporation, as well as the entropy production $\Delta_{\rm PBH}=\tilde{S}_{2}/\tilde{S}_{1}$, where $\tilde{S}_{i}\propto a_{i}^3/z_{i}^3$, is the ratio between the total entropy before ($i=1$) and after ($i=2$) the PBH evaporation. 
\begin{figure}
\centering
\includegraphics[width=0.478\textwidth]{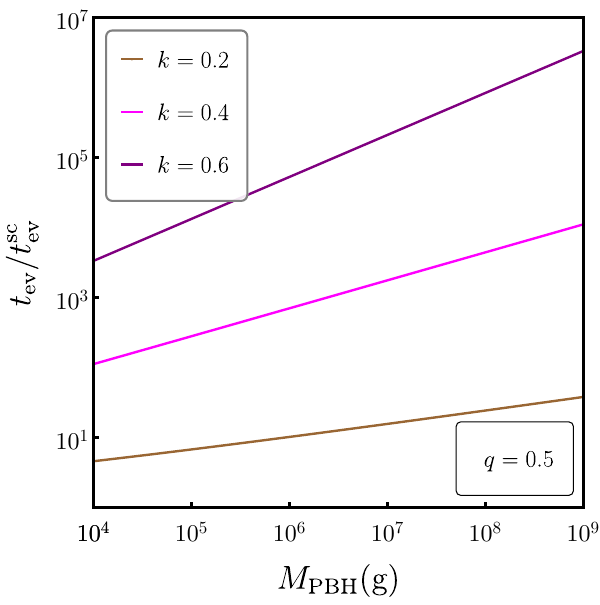}
\includegraphics[width=0.49\textwidth]{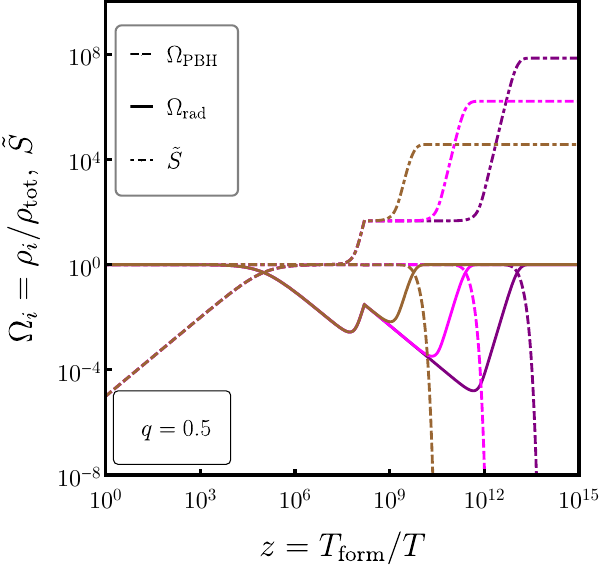}
\caption{\textbf{Left:} PBH lifetime as a function of $M_{\rm PBH}$ for different values of $k$, with $q$ fixed at 0.5. \textbf{Right:} The evolution of normalized energy densities for PBHs $\Omega_{\rm PBH} =\rho_{\rm PBH}/(\rho_{\rm PBH}+\rho_{\rm rad})$ and radiation $\Omega_{\rm rad}=\rho_{\rm rad}/(\rho_{\rm PBH}+\rho_{\rm rad})$, and the entropy $\tilde{S}$ as a function of the auxiliary variable $z = T_{\rm form}/T$. The different colors correspond to the benchmarks $k= 0.2$ (brown), $0.4$ (magenta), and $0.6$ (purple) with fixed $q = 0.5$.}\label{fig1}
\end{figure}

From the inspection of the left panel of Fig.~\ref{fig1}, it is clear that the memory burden significantly extends the lifespan of PBHs, with the effect becoming more pronounced as $k$ increases for a fixed $q=0.5$. This extension of the PBH lifespan results in a prolonged period of early matter domination provided by the PBHs, which leads to substantial entropy injection into the thermal bath. The right panel, which shows the dynamical evolution of energy densities, clearly illustrates the timeline of PBH evolution in the early universe. Before PBH formation, the universe was in a state of radiation domination, with $\Omega_{\rm rad}\approx 1$. After PBHs form and as the universe expands, the differing scaling behaviors of the radiation energy density $\rho_{\rm rad}$ and the PBH energy density $\rho_{\rm PBH}$ lead to a transition to matter domination, as indicated by $\Omega_{\rm PBH}$ (dashed lines) increasing towards 1, while $\Omega_{\rm rad}$ (solid lines) decreases. However, similar to the standard scenario without the memory burden effect, as PBHs lose mass, their evaporation rate increases due to the inverse dependence on $M_{\rm PBH}$ in Eq.~\eqref{eq1}, eventually reaching a turning point at 
$x_{\rm mb}=T_{\rm form}/T_{\rm mb}$, where $\Omega_{\rm rad}$ begins to rise again. In contrast, due to the memory burden effect, once the PBH mass reaches a threshold value $M_\text{mb} = q M_\text{\rm in}$ and the memory burden effect sets in, $\Omega_{\rm rad}$ begins to decrease once more. Thus, the key impact of the memory burden effect is to delay the return to radiation domination, effectively prolonging the period of matter domination.\footnote{In principle, for sufficiently small values of 
$q\,(\ll 0.5)$, another intermediate radiation domination phase could arise starting around $T_{\rm mb}$ (see Appendix~\ref{apa}). While our primary results focus on $q=0.5$, we will show that given a uniform prior for 
$q\in[0.2,0.8]$, the PTA fit is not so sensitive to $q$. However, for smaller values of $q$, the GW spectrum exhibits a kink-like feature exclusive to memory-burdened PBHs.} The ``second phase" of matter domination, which begins around $x_{\rm mb}$, is influenced by the residual PBH mass $M_{\rm mb}$ and is much more sensitive to $k$.

Given this post-inflationary cosmological set-up, i.e., the EMD induced by the memory-burdened PBHs, we will now provide a brief overview of BGWs and explore how they may encode information about the properties of these PBHs.

\section{Imprint of modified PBH evaporation dynamics on BGWs from inflation \label{sec:bgw}}

In this section, we will provide a brief overview of the production of GWs during inflation and how they travel through different cosmic eras until they arrive in the present day. GWs are described as a perturbation in the FLRW line element: 
\begin{equation}
ds^2=a(\tau)\left[-d\tau^2+(\delta_{ij}+h_{ij})dx^idx^j)\right]\,,
\end{equation}
with $\tau$ being the conformal time, and $a(\tau)$ the scale factor. The GWs are interpreted by the transverse and traceless components of the $3\times 3$ symmetric matrix $h_{ij}$, satisfying the conditions $\partial_ih^{ij}=0$, and $\delta^{ij}h_{ij}=0$. Since the GWs are so feeble, i.e. $|h_{ij}|\ll1$, the linearized evolution equation 
\begin{equation}
\partial_\mu(\sqrt{-g}\partial^\mu h_{ij})=16\pi a^2(\tau) \mathcal{\pi}_{ij}\label{lineq}
\end{equation}
is adequate for studying their propagation. The tensor part of the anisotropy stress, $\pi_{ij}$, coupled with $h_{ij}$, acts as an external source. It is useful to express $h_{ij}$ in the Fourier space as
\begin{equation}
h_{ij}(\tau, \vec{x})=\sum_\lambda\int \frac{{\rm d}^3\vec{k}}{(2\pi)^{3/2}} e^{i\vec{k}.\vec{x}}\epsilon_{ij}^\lambda(\vec{k})h_{\vec{k}}^\lambda(\tau)\,,\label{fug}
\end{equation}
where the index $\lambda=``+/-"$ represents the two polarisation states of GWs. The polarisation tensors, besides being transverse and traceless, also fulfil the conditions
\begin{equation}
\epsilon^{(\lambda)ij}(\vec{k})\epsilon_{ij}^{(\lambda^\prime)}(\vec{k})=2\delta_{\lambda\lambda^\prime} \quad {\rm and} \quad \epsilon^{(\lambda)}_{ij}(-\vec{k})=\epsilon^{(\lambda)}_{ij}(\vec{k})\,.
\end{equation}
Assuming that each polarisation state evolves identically and isotropically, we can simplify the notation by letting $h_{\vec{k}}^\lambda(\tau)$ as $h_{k}(\tau)$, where $k=|\vec{k}|=2\pi f$ with $f$ being the frequency of the GWs today at $a_0=1$. Considering the sub-dominant contribution from $\pi_{ij}$, the equation governing the propagation of GWs in Fourier space can be expressed as
\begin{equation}
\ddot{h}_k+2\frac{\dot{a}}{a}\dot{h}_k+k^2h_k=0\,, \label{prpeq}
\end{equation}
where the dot represents a derivative with respect to the conformal time. By utilizing Eq.s~\eqref{fug} and~\eqref{prpeq}, we can compute the energy density of the GWs as~\cite{WMAP:2006rnx}
\begin{equation}
\rho_{\rm GW}=\frac{1}{32\pi G}\int\frac{{\rm d}k}{k}\left(\frac{k}{a}\right)^2T_T^2(\tau, k)P_T(k)\,,\label{gw1}
\end{equation}
where $T_T^2(\tau, k)=|h_k(\tau)|^2/|h_k(\tau_i)|^2$ represents a transfer function derived from Eq.~\eqref{prpeq}, with $\tau_i$ denoting an initial conformal time. The primordial power spectrum, $P_T(k)=k^3|h_k(\tau_i)|^2 / \pi^2$ is associated with inflation models of specific forms and is parametrised as a power-law, given by
\begin{equation}
P_T(k)=r A_s(k_*)\left(\frac{k}{k_*}\right)^{n_T}\,,\label{ps}
\end{equation}
where $r\lesssim 0.06$ \cite{BICEP2:2018kqh} is the tensor-to-scalar-ratio,  $A_s \simeq 2\times 10^{-9}$ represents the amplitude of scalar perturbation at the pivot scale $k_*=0.01~{\rm  Mpc^{-1}}$, and $n_T$ represents the tensor spectral index. Interestingly, the standard slow-roll inflation models adhere to a consistency relation $n_T=-r/8$ \cite{Liddle:1993fq}, leading to GWs that are slightly red-tilted with $n_T\lesssim 0$. In contrast, we have considered GWs with a significant blue tilt, where $n_T>0$, and we have assumed this value to remain constant throughout. The GW energy density, which is crucial for detection purposes, can be expressed as 
\begin{equation}
\Omega_{\rm GW}(k)=\frac{k}{\rho_c}\frac{{\rm d}\rho_{\rm GW}}{{\rm d}k}\,,
\end{equation}
with $\rho_c=3H_0^2/8\pi G$. From Eq.~\eqref{gw1}, the quantity $\Omega_{\rm GW}(k)$ is computed as 
\begin{equation}
\Omega_{\rm GW}(k)=\frac{1}{12H_0^2}\left(\frac{k}{a_0}\right)^2T_T^2(\tau_0,k)P_T(k)\quad \text{with}~~\tau_0=1.4\times 10^4 {\rm ~Mpc}\,.\label{GWeq}
\end{equation}
In the presence of an intermediate matter domination (here by the memory-burdened PBHs), the transfer function $T_T^2(\tau_0,k)$ is given by~\cite{t1,t2,t3,t4,t5,t6}
\begin{equation}
T_T^2(\tau_0,k)= F(k)T_1^2(\zeta_{\rm eq})T_2^2(\zeta_{\text{ev}})T_3^2(\zeta_{\rm dom})T_2^2(\zeta_{R})\,, \label{TT}
\end{equation}
where the quantity $F(k)$ takes the expression
\begin{equation}
F(k) = \Omega_m^2\left( \frac{g_*(T_{k,\rm in})}{g_{*0}}\right)\left(\frac{g_{*0s}}{g_{*s}(T_{k,\rm in})}\right)^{4/3}\left(\frac{3j_1(k\tau_0)}{k\tau_0}\right)^2 \,. \label{fk}
\end{equation}
Here, $T_{k,\rm in}$ is the temperature corresponding to the horizon entry of the $k$th mode, $j_1(k\tau_0)$ is the spherical Bessel function, $g_{*0}=3.36$, $g_{*0s}=3.91$, and an approximate form of the scale-dependent $g_*$ is given by~\cite{gs1,gs2,t6}
\begin{equation}
g_{*(s)}(T_{k,\rm in})=g_{*0(s)}\left(\frac{A+{\tanh k_1}}{A+1}\right)\left(\frac{B+{\tanh k_2}}{B+1}\right)\,,
\end{equation}
where 
\begin{equation}
A=\frac{-1-10.75/g_{*0(s)}}{-1+10.75/g_{*0(s)}}\,,~~B=\frac{-1-g_{\rm max}/10.75 }{-1+g_{\rm max}/10.75}\,, 
\end{equation}
and
\begin{equation}
k_1=-2.5~\log_{10}\left(\frac{k/2\pi}{2.5\times 10^{-12}~{\rm Hz}}\right)\,, ~~k_2=-2.0~{\log}_{10}\left(\frac{k/2\pi}{6.0\times 10^{-9}~{\rm Hz}}\right)\,,
\end{equation}
with $g_{\rm max}$ being $\simeq 106.75$. The individual transfer functions read as
\begin{eqnarray}
    T_1^2(\zeta) &=& 1+1.57\,\zeta+ 3.42 \,\zeta^2\,,\\
    T_2^2(\zeta) &=& \left(1-0.22\,\zeta^{1.5}+0.65\,\zeta^2 \right)^{-1}\,,\\
    T_3^2(\zeta) &=& 1+0.59\,\zeta+0.65 \, \zeta^2\,,
\end{eqnarray}
with $\zeta_i =  k/k_i$. The modes $k_i$ given by
\begin{eqnarray}
    k_{\rm eq} &=& 7.1\times 10^{-2}\,\Omega_m h^2\,{\rm Mpc^{-1}}\,,\\
    k_{\text{ev}} &=& 1.7\times 10^{14}\left(\frac{g_{*s}(T_{\rm dec})}{106.75}\right)^{1/6}\left(\frac{T_{\text{ev}}}{10^7 \rm GeV}\right)\,{\rm Mpc^{-1}} \,,\label{flow1} \\
    k_{\text{dom}} &=& 1.7\times 10^{14} \Delta_{\rm PBH}^{2/3}\left(\frac{g_{*s}(T_{\rm ev})}{106.75}\right)^{1/6}\left(\frac{T_{\rm ev}}{10^7 \rm GeV}\right)\,{\rm Mpc^{-1}}\,, \label{fdip1}\\
    k_{R} &=&1.7\times 10^{14}\Delta_{\rm PBH}^{-1/3}\left(\frac{g_{*s}(T_{R})}{106.75}\right)^{1/6}\left(\frac{T_{R}}{10^7 \rm GeV}\right)\,{\rm Mpc^{-1}}\label{fhigh} \,,
\end{eqnarray}
re-enter the horizon at $T_{\rm eq}$ (the standard matter-radiation equality temperature), at $T_{\rm ev}$ (final evaporation temperature of memory-burdened PBHs), at $T_{\rm dom}$ (the temperature at which the PBH start to dominate the energy density) and at  $T_{R}$ (when the universe reheats after inflation), respectively. The last three temperatures correspond to three characteristic frequencies describing a double peak GW signal spanning the nHz-Hz range (see, e.g. Fig.~\ref{fig3}). The explicit expressions for those frequencies can be derived as 
\begin{equation}
f_{\rm low}^{\rm peak}\simeq 220.8 {\:\rm nHz}\,\mathcal{T}(q,k)^{-1/2}\left( \frac{g_{\rm *s}(T_{\rm ev})}{106.75}\right)^{1/6}\left( \frac{g_{\rm *}}{100}\right)^{1/4}\left( \frac{10^7~{\rm g}}{M_{\rm PBH}}\right)^{3/2}\,,\label{flow2} 
\end{equation}
\begin{equation}
f_{\rm dip}\simeq 1.426\times 10^{-4} {\:\rm Hz} \,\mathcal{T}(q,k)^{-1/5}\left( \frac{\beta}{10^{-8}}\right)^{2/3}\left( \frac{g_{\rm *}}{100}\right)^{-1/4}\left( \frac{g_{\rm *s}(T_{\rm dom})}{106.75}\right)^{1/6}\left(\frac{10^7~{\rm g}}{M_{\rm PBH}} \right)^{5/6}\,,\label{fdip2} 
\end{equation}
\begin{equation}
f_{\rm high}^{\rm peak}\simeq 1.426\times 10^{4} {\:\rm Hz}\,\mathcal{T}(q,k)^{-1/6}\left( \frac{\beta}{10^{-8}}\right)^{-1/3}\left( \frac{g_{\rm *}}{100}\right)^{-1/12}\left( \frac{g_{\rm *s}(T_{\rm R})}{106.75}\right)^{1/6}\left(\frac{10^7~{\rm g}}{M_{\rm PBH} }\right)^{5/6}\,,\label{fhigh}
\end{equation}
where $\mathcal{T}(q,k)=(1-q^3)+\frac{3}{3+2k} q^{3+2k}S^{k}(M_{\rm PBH})= t_{\rm ev}/t_{\rm ev}^{\rm sc}$. In the present paper, we consider the following values for the cosmological parameters: $H_0\simeq 2.2 \times 10^{-4}~\rm Mpc^{-1}$, $\Omega_m = 0.31$, and $h=0.7$. 

Before we proceed to the numerical discussion, let us point out that the quantity $\Omega_{\rm GW}(k)h^2$ is constrained by two robust bounds on the SGWB placed by the effective number of relativistic species during BBN~\cite{Peimbert:2016bdg} and by the LIGO measurements on SGWB~\cite{LIGOScientific:2016jlg}. The BBN constraint reads
\begin{equation}
\int_{f_{\rm low}}^{f_{\rm high}} {\rm d} f ~f^{-1}\Omega_{\rm GW}(f)h^2\lesssim 5.6\times 10^{-6}\Delta N_{\rm eff}\,,
\end{equation}
with $\Delta N_{\rm eff}\lesssim 0.2$. The lower limit of the integration is the frequency that represents the scale entering the horizon at the BBN epoch, which we take as $f_{\rm low}\simeq 10^{-10}$~Hz. We consider the upper limit $f_{\rm high}\simeq 10^{7}$~Hz corresponding to $T_R\equiv T_R^{\rm max}\simeq 5 \times 10^{15}$ GeV. For the LIGO constraint, we rely on a crude estimation by considering a pivot frequency $f_{\rm LIGO}=25~{\rm Hz}$ and discarding the GWs having an amplitude more than $8.33\times 10^{-9}$ at $f_{\rm LIGO}$~\cite{KAGRA:2021kbb}.

\section{Results \label{sec:results}}

%
%
\begin{table}
\begin{center}
\begin{tabular}{|B{2.5cm}||B{2.4cm}|B{2.7cm}|B{2.4cm}|B{2.55cm}|}
\hline
\multicolumn{5}{|c|}{\textbf{NANOGrav-15yrs}} \\
\hline
Parameters & $n_T$  & $\log_{10}(r)$ & $\log_{10}({M_{\rm PBH}/\rm g})$ & $k$\\
\hline
Priors & Uniform$\left[ 0,5\right]$  & Uniform$\left[-20, 0\right]$  & Uniform$\left[ 0,9\right]$  & Uniform$\left[ 0,3\right]$  \\
\hline
Best fit & $1.8961\pm 0.4159$  & $-9.5074\pm 3.4893$ & $3.3261\pm 2.1892$ & $0.7461\pm 0.6014$\\
\hline
\hline
\multicolumn{5}{|c|}{\textbf{IPTA Data Release 2}} \\
\hline
Parameters & $n_T$  & $\log_{10}(r)$ & $\log_{10}({M_{\rm PBH}/\rm g})$& $k$\\
\hline
Priors & Uniform$\left[ 0,5\right]$  & Uniform$\left[-20, 0\right]$  & Uniform$\left[ 0,9\right]$  & Uniform$\left[ 0,3\right]$  \\
\hline
Best fit & $1.5337\pm 0.5512$  & $-6.3433\pm 4.4045$ & $3.1780\pm 2.2269$ & $0.8278\pm 0.6089$\\
\hline
\end{tabular}
\end{center}
\caption{The range of uniform priors used in the posterior plots in Fig.~\ref{fig2}, with their best-fit values with maximum likelihood, taking $q=0.5$. The impact of the other parameters $\beta$ and $T_R$ on the Bayesian analysis is almost negligible due to the low-frequency range of PTA data.}\label{t1}
\end{table}
\begin{figure}
\centering

\includegraphics[width=0.9\textwidth]{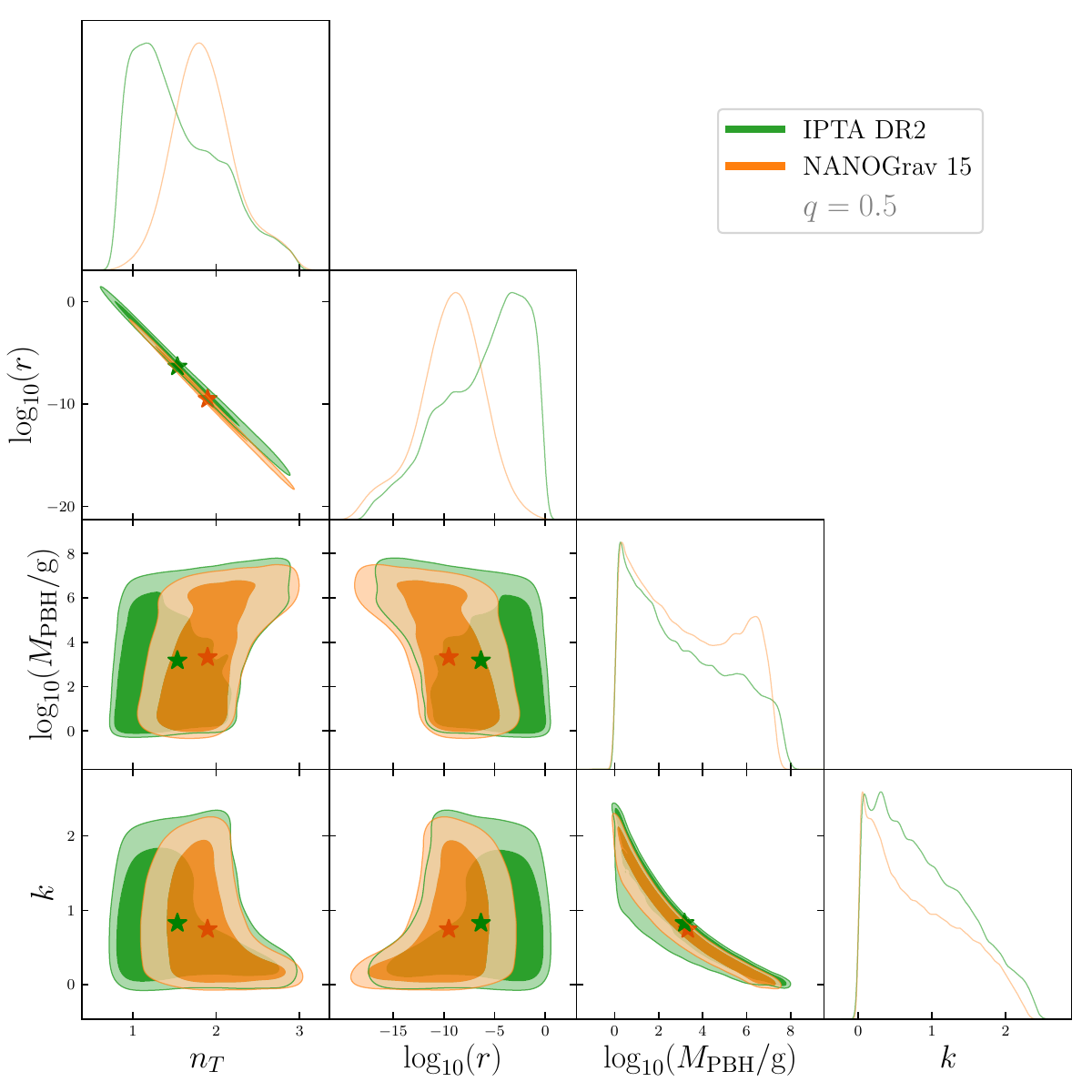}
\caption{Marginalized 2D posterior distributions of the memory-burdened PBH and inflationary parameters with $q=0.5$. The green (orange) color corresponds to IPTA-DR2 (NANOGrav-15yrs) data, while the different shadings refer to $1\sigma$ and $2\sigma$ regions.}\label{fig2}
\end{figure}
\begin{figure}
\centering
\includegraphics[width=0.49\textwidth]{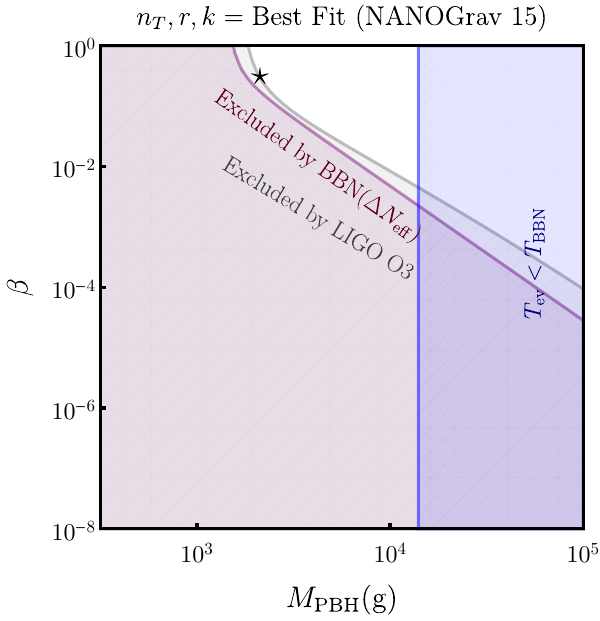}
\includegraphics[width=0.49\textwidth]{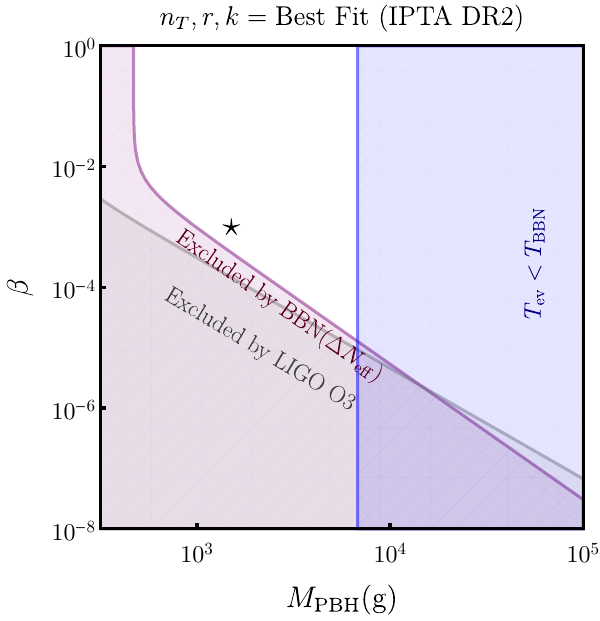}
\includegraphics[width=0.49\textwidth]{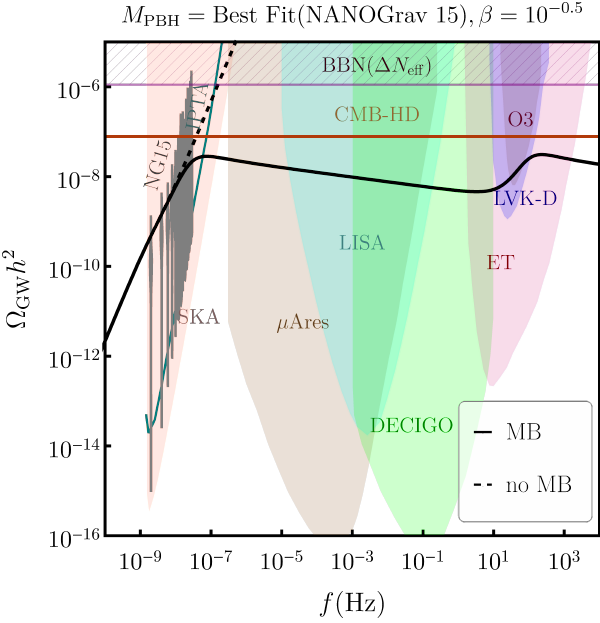}
\includegraphics[width=0.49\textwidth]{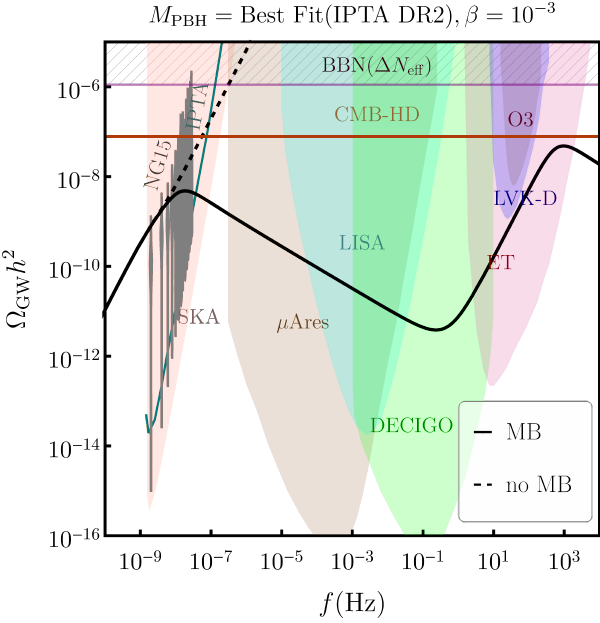}
\caption{\textbf{Top panels:} Exclusion regions in the $\beta-M_{\rm PBH}$ parameter space, as determined by BBN (purple) and LIGO-O3 (gray),  for the NANOGrav-15yrs (left panel) and IPTA-DR2 (right panel) best-fit values of $n_T$,  $r$, and  $k$ (see Tab.~\ref{t1}). The symbol ``$\star$'' indicates the best-fit value of $ M_{\rm PBH}$ from NANOGrav-15yrs at $\beta = 10^{-0.5}$ and IPTA-DR2 at $\beta = 10^{-3}$.
\textbf{Bottom panels:} The corresponding BGW spectrum for the benchmark cases as mentioned above is represented by solid lines for the case with memory burden and dashed lines for the case without memory burden.}
\label{fig3}
\end{figure}

Here we identify the region of parameter space that optimally fits the PTA data. We employ Bayesian statistics to analyze PTA data and their posterior best-fit values with maximum likelihood, including the IPTA-DR2 and NANOGrav-15yrs datasets, utilizing the publicly available code \texttt{PTArcade}~\cite{Mitridate:2023oar} in the \texttt{ceffyl}~\cite{Lamb:2023jls} mode. To mitigate the pulsar's intrinsic excess noise at high frequencies~\cite{NANOGrav:2023gor, Antoniadis:2022pcn}, we include only the first 13 and 14 frequency components from the IPTA-DR2 and NANOGrav-15yrs datasets, respectively. In Fig.~\ref{fig2} ($q=0.5$) and Fig.~\ref{fig4} ($q\in[0.2,0.8]$), we have illustrated the resulting marginalized 2D posterior distributions: green for IPTA-DR2 and orange for NANOGrav-15yrs, using the \texttt{GetDist}~\cite{Lewis:2019xzd}. A few remarks on PTA fits are in order. The PTA fit shows limited sensitivity to $\beta$ and the reheating temperature $T_R$. Consideration of their appropriate values nonetheless remains essential to produce GW spectrum at higher frequencies consistent with LIGO-O3 bound and BBN constraint on $\Delta N_{\rm eff}$, where the latter is saturated by the GW amplitude at $f_{\rm high}^{\rm peak}\sim T_R$. To assess the sensitivity of the memory-burdened PBH parameters to PTA data, it is sufficient to impose the constraints $t_{\rm ev} \leq t_{\rm BBN}$ and $h^2\Omega_{\rm GW}(f_{\rm low}^{\rm peak})< 5.6\times 10^{-6}\Delta N_{\rm eff}$. With these considerations, we now provide detailed discussions for two cases: 
first, fixing $q=0.5$, which is the standard assumption in the literature on memory burden effects in black holes, and secondly, varying $q$ within $[0.2,0.8]$ to see how it impacts the results.
\subsection{The case with $q=0.5$}

For $q=0.5$, we consider the uniform priors, shown in Tab.~\ref{t1}, in our MCMC analysis. The set consists of 4 parameters: ($n_T$, $r$) describing the blue-tiled SGWB spectrum from inflation, and the memory-burdened PBH parameters ($M_{\rm PBH}$, $k$) impacting the SGWB at low frequencies relevant to PTA. Some important points are evident from the posterior distributions of Fig.~\ref{fig2}. We find a strong correlation between the spectral index $n_T$ and the tensor-to-scalar ratio $r$, which is consistent with the $n_T = -0.14 \log_{10} r + 0.58$ fit reported by NANOGrav \cite{NANOGrav:2023hvm}. This correlation implies an approximate reflection symmetry between $n_T$ {\it vs} ($M_{\rm PBH},k$) and $r$ {\it vs} ($M_{\rm PBH},k$), which is indeed evident in Fig.~\ref{fig2}.

To understand the fits in more detail, consider first the simple case of fitting all the NANOGrav frequency bins with a simple power-law $\Omega_{\rm GW}\propto f^{n_T}$.  This results in $n_T\in[0.6,3]$ at 2$\sigma$ with the best-fit value $n_T\simeq 2$ \cite{NANOGrav:2023hvm}.  The upper (lower) limit on $n_T$ simply comes from the fact that the amplitude increases too quickly (slowly) to pass through enough of the frequency bins from NANOGrav.  If instead of a simple power law, there is a break at some frequency, with the amplitude increasing less rapidly for higher frequencies, then it may be possible for the GW signal to fit the NANOGrav data for $n_T > 3$.  


EMD provided by the memory-burdened PBHs can provide a break in the power law for frequencies corresponding to PBH evaporation, $f\sim f_{\rm low}^{\rm peak}$. More precisely, at $f_{\rm low}^{\rm peak}$, the BGWs undergo a spectral turnover from a frequency scaling of $f^{n_T}$ (for $f<f_{\rm low}^{\rm peak}$) to $f^{n_T - 2}$ (for $f>f_{\rm low}^{\rm peak}$). At the same time, memory-burdened PBHs with sufficiently large $M_{\rm PBH}$ or large $k$ exhibit significantly longer lifetimes, shifting $f_{\rm low}^{\rm peak}$ towards much lower frequencies within the PTA band.  One might expect an early enough break in the spectrum could then allow for a good fit of the NANOGrav data for $n_T>3$, with the GW spectrum being less steep for $f>f_{\rm low}^{\rm peak}$. However we find that even if $T_{\rm ev}\simeq T_{\rm BBN}$, which corresponds to the minimum allowed value of $f_{\rm low}^{\rm peak}$, this frequency is too high to allow the GW spectrum to fit the NANOGrav data for $f>f_{\rm low}^{\rm peak}$. Instead, the GW spectrum remains too steep and the amplitude overshoots the PTA signal, so the $2\sigma$ upper limit $n_T\simeq 3$ does not change significantly compared to the standard power-law fit. 

On the other hand, for smaller spectral indices, e.g., $n_T<1$ (which were allowed within 2$\sigma$ in a pure power-law fit), the spectral turnover at $f_{\rm low}^{\rm peak}$ happens at significantly lower frequencies within the PTA frequency band, with the lowest value of $f_{\rm low}^{\rm peak}$ corresponding to, once again, $T_{\rm ev}\simeq T_{\rm BBN}$. Thus, the amplitude increases more slowly, giving a worse fit for lower $n_T$ values and eliminating $n_T<1$ values from the 2$\sigma$ interval. A similar spectral turnover within the PTA band occurs also for the intermediate spectral index value $n_T\simeq 2$, such that large $M_{\rm PBH}$ or $k$ can fit the PTA data only at 2$\sigma$ for this value of the spectral index. However, for moderate values of $M_{\rm PBH}$ and $k$, the turnover frequency $f_{\rm low}^{\rm peak}$ is larger than the highest frequency bin in the PTA band. Thus, values of $M_{\rm PBH}\sim10^{3.2}$ g, $k\sim 0.8$ when $n_T\sim 2$ can give a good fit to the PTA data, providing the best-fit values for NANOGrav, with the best-fit point for IPTA data being similar.

Note that standard PBHs without memory burden effect~\cite{Datta:2023xpr} are obtained in the limit $k\rightarrow 0$. For comparison, we show this fit in Appendix~\ref{apb}, where the fit favors slightly higher values of $n_T$ and lower $r$ and very large PBH masses are needed to increase the lifetime sufficiently.  The memory burden effect with non-zero $k$ also increases the lifetime, allowing longer lifetimes for smaller PBH masses and extending the range of PBH masses that can fit the PTA data down to around $1$ g \footnote{This lower bound $M_{\rm PBH}\gtrsim 1$g arises from the maximum possible reheating temperature $T_{R}^{\rm max}\sim 5\times 10^{15}$ GeV.}.  Thus, reducing the PBH masses can be compensated by increasing $k$, leading to the rough $k\sim 1/M_\text{PBH}$ correlation seen in the bottom row, third panel of  Fig.~\ref{fig2}.   



While it seems from Fig.~\ref{fig2} that the PTA data fits allow for a large range of PBH masses, it does not account for the BBN constraint on $\Delta N_{\rm eff}$ and also LIGO-O3 bounds on higher frequencies. As mentioned previously, to describe a viable GW spectrum across a wide frequency range, the parameters $\beta$ and $T_R$ are crucial. In the top panels (left: NANOGrav-15yrs, right: IPTA-DR2) of Fig.~\ref{fig3}, we show the allowed regions on the $M_{\rm PBH}$-$\beta$ plane for the minimally allowed value of $T_R\simeq T_{\rm form}$. The white region is consistent with PTA data as well as BBN and LIGO-O3 bounds. These plots show that a viable, wide-spanning GW signal can only be obtained for a constrained range of PBH mass. It is, however, remarkable that even PBHs as light as $10^2$ g can leave distinct imprints on a strong, wide-spanning GW signal owing to the memory burden effect. In the bottom panels, we show the GW spectrum for the best-fit values. We have chosen $\beta=10^{-0.5}$ and $\beta=10^{-3}$ for NANOGrav-15yrs and IPTA-DR2, respectively.

\subsection{The case with varying $q$}

%
%
\begin{table}
\begin{center}
\begin{tabular}{|B{2.5cm}||B{2.4cm}|B{2.7cm}|B{2.4cm}|B{2.55cm}|B{2.4cm}|}
\hline
\multicolumn{6}{|c|}{\textbf{NANOGrav-15yrs}} \\
\hline
Parameters & $n_T$  & $\log_{10}(r)$ & $\log_{10}({M_{\rm PBH}/\rm g})$& $q$ & $k$\\
\hline
Priors & Uniform$\left[ 0,5\right]$  & Uniform$\left[ 0,-20\right]$  & Uniform$\left[ 0,9\right]$  & Uniform$\left[ 0.2,0.8\right]$  & Uniform$\left[ 0,3\right]$  \\
\hline
Best fit & $1.8999\pm 0.4092$  & $-9.5293\pm 3.4301$ & $3.4046\pm 2.1905$ & $0.4662\pm 0.1713$ & $0.7632\pm 0.6188$\\
\hline
\hline
\multicolumn{6}{|c|}{\textbf{IPTA Data Release 2}} \\
\hline
Parameters & $n_T$  & $\log_{10}(r)$ & $\log_{10}({M_{\rm PBH}/\rm g})$& $q$ & $k$\\
\hline
Priors & Uniform$\left[ 0,5\right]$  & Uniform$\left[ 0,-20\right]$  & Uniform$\left[ 0,9\right]$  & Uniform$\left[ 0.2,0.8\right]$  & Uniform$\left[ 0,3\right]$  \\
\hline
Best fit & $1.5184\pm 0.5432$  & $-6.1650\pm 4.3346$ & $3.1728\pm 2.2335$ & $0.4805\pm 0.1712$ & $0.8486\pm 0.6262$\\
\hline
\end{tabular}
\end{center}
\caption{The range of uniform priors used in the posterior plots in Fig.~\ref{fig4} with their best fits with maximum likelihood.}\label{t2}
\end{table}

\begin{figure}
\centering
\includegraphics[width=0.9\textwidth]{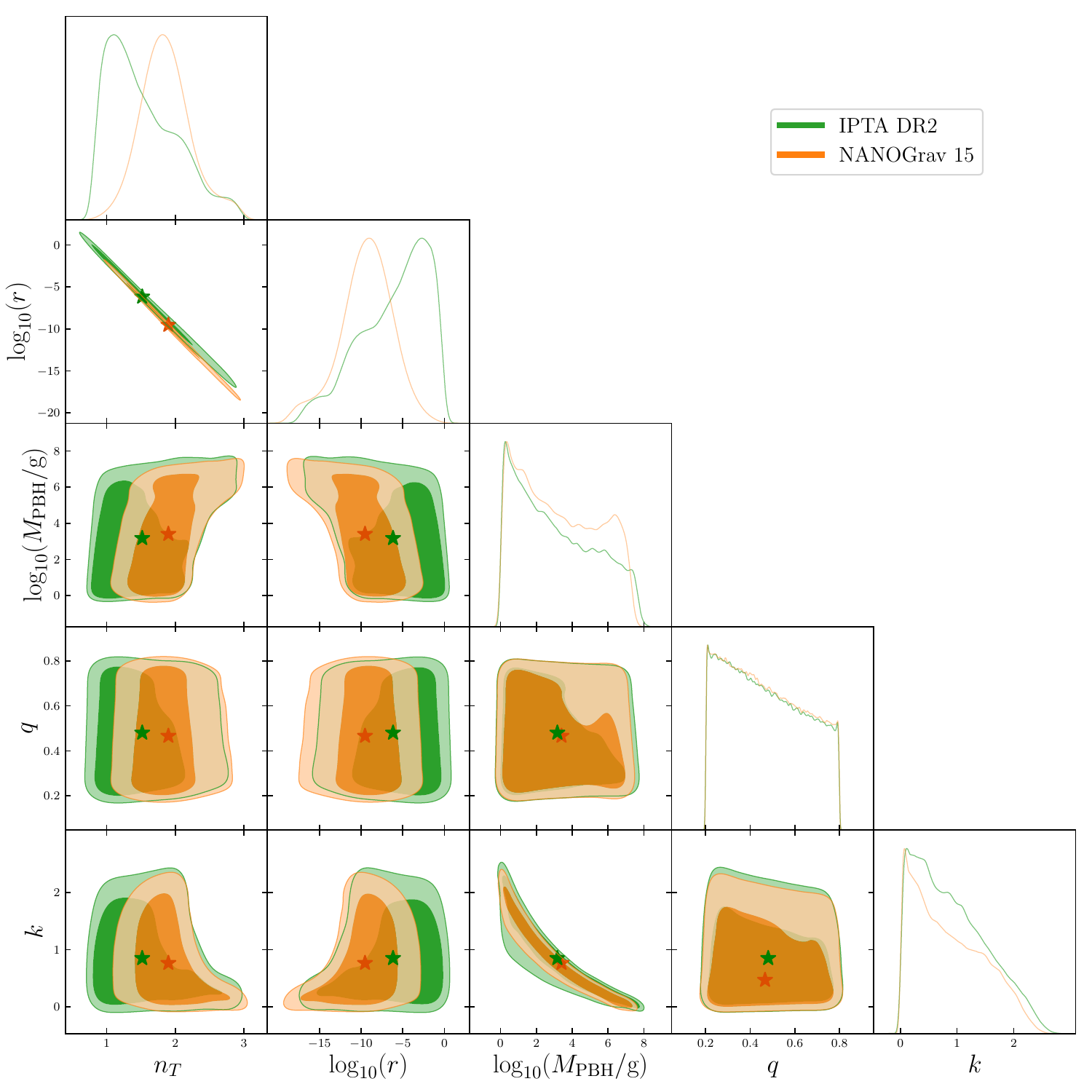}
\caption{Marginalized 2D posterior distributions of the memory-burdened PBH and inflationary parameters with varying $q$ in $[0.2,\,0.8]$. The green (orange) color corresponds to IPTA-DR2 (NANOGrav-15yrs) data, while the different shadings refer to $1\sigma$ and $2\sigma$ regions.}\label{fig4}
\end{figure}
\begin{figure}
\centering
\includegraphics[width=0.49\textwidth]{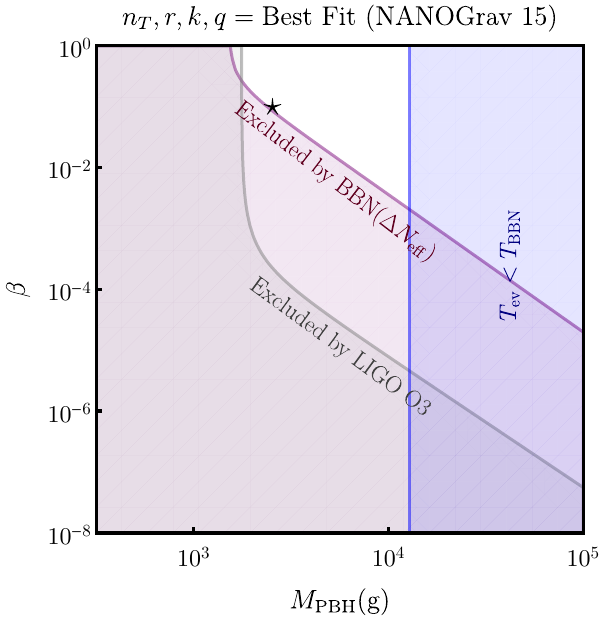}
\includegraphics[width=0.49\textwidth]{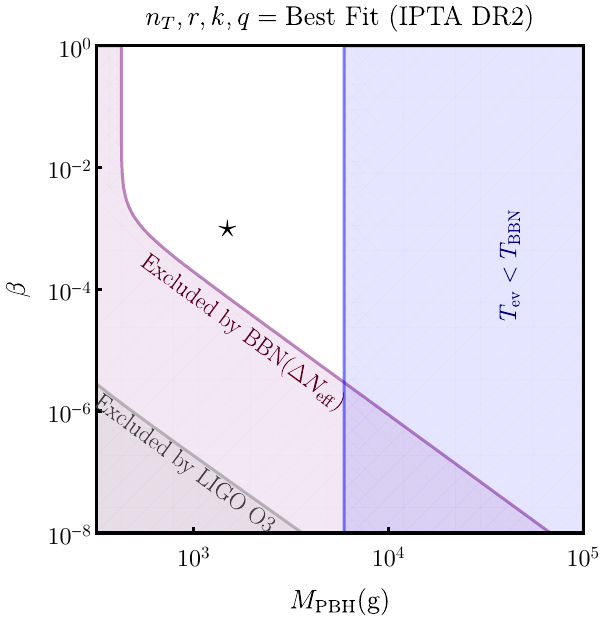}\\
\includegraphics[width=0.49\textwidth]{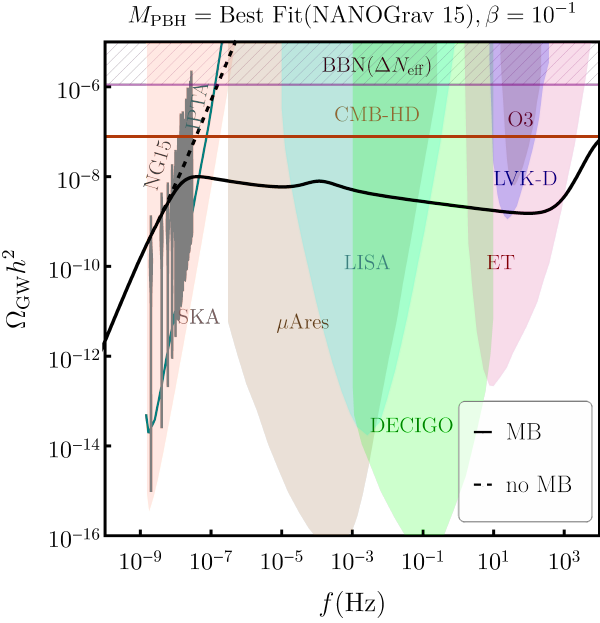}
\includegraphics[width=0.49\textwidth]{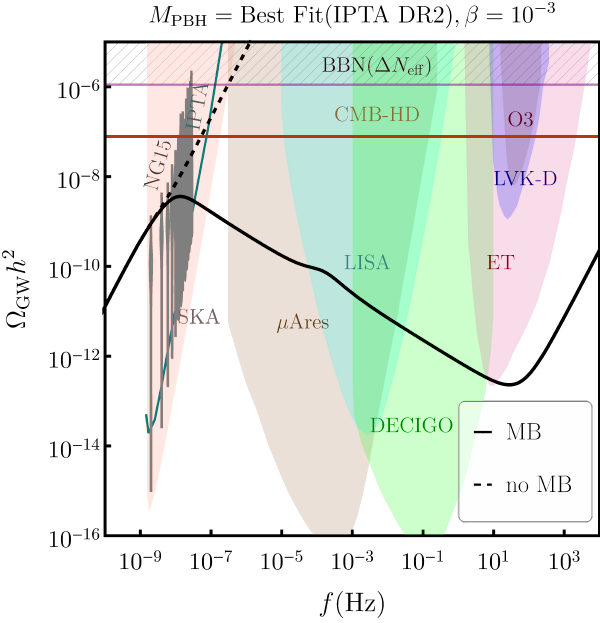}
\caption{\textbf{Top panels:} Exclusion regions in the $\beta-M_{\rm PBH}$ parameter space, as determined by BBN (purple) and LIGO-O3 (gray),  for the NANOGrav-15yrs (left panel) and IPTA-DR2 (right panel) best-fit values of $n_T$,  $r$, $k$ and $q$ (see Tab.~\ref{t2}). The symbol ``$\star$'' indicates the best-fit value of $ M_{\rm PBH}$ from NANOGrav-15yrs at $\beta = 10^{-1}$ and IPTA-DR2 at $\beta = 10^{-3}$.
\textbf{Bottom panels:} The corresponding BGW spectrum for the benchmark cases as mentioned above is represented by solid lines for the case with memory burden and dashed lines for the case without memory burden. Both the plots (with $q<0.5$) also highlight the potential for complementary detection of the characteristic kink associated with the memory burden effect.}
\label{fig5}
\end{figure}

We now turn our attention towards another interesting possibility with variable $q$. For phenomenological purposes, we vary it from 0.2 to 0.8, thus having an additional handle over the interplay of PBH domination and PTA data. The uniform priors employed in the MCMC analysis, with the additional parameter $q$, are reported in Tab.~\ref{t2}. Some important points are clear from the posterior distributions shown in Fig.~\ref{fig4}. First, there exists a very mild positive correlation between $q$ and the inflationary parameter $r$, and consequently a negative correlation between $q$ and $n_T$. The reason is the same as described in the previous case: as $q$ increases, the residual energy density after semiclassical evaporation rises, leading to a prolonged period of PBH domination. While $q$ does influence inflationary observables, its impact is less pronounced than that of the parameters $k$ or $M_{\rm PBH}$. Consequently, we observe a modest effect on $n_T$ and $r$ when analysing the PTA data. Furthermore, increasing $q$ leads to a longer PBH lifetime, causing heavier PBHs to be excluded by the $t_{\rm ev} \leq t_{\rm BBN}$ bound. Conversely, while a scenario with a lower $q$ with our chosen prior range for $k$ allows for significantly higher values, increasing $q$ can lead to conflicts with a larger $k$. This arises because both $k$ and $q$ influence the PBH lifetime in a similar manner, resulting in constraints imposed by the $t_{\rm ev} \leq t_{\rm BBN}$ at their extreme values.

The description of Fig.~\ref{fig5} remains the same as that of Fig.~\ref{fig3}, apart from the kink in the middle of the GW spectrum. This happens owing to an intermediate radiation domination for $q<0.5$. This characteristic kink feature, which is unique to memory-burdened PBHs, is described analytically as 
\begin{equation}
f_{\rm mb}^{\rm kink}\simeq 220.8 {\:\rm nHz}\mathcal{T}(q,k)^{-1/2}\left(\frac{(1-q)^{1/4}}{q(1-q^3)}\right)^{1/3}\left( \frac{g_{\rm *s}(T_{\rm ev})}{106.75}\right)^{1/6}\left( \frac{g_{\rm *B}}{100}\right)^{1/4}\left( \frac{10^7 g}{M_{\rm PBH}}\right)^{3/2}\, , \label{fdip3} 
\end{equation}
and should be present when $q< 0.5$, provided that the semiclassical computation of the PBH lifetime remains valid in this regime. 

\begin{figure}
\centering
\includegraphics[width=0.49\textwidth]{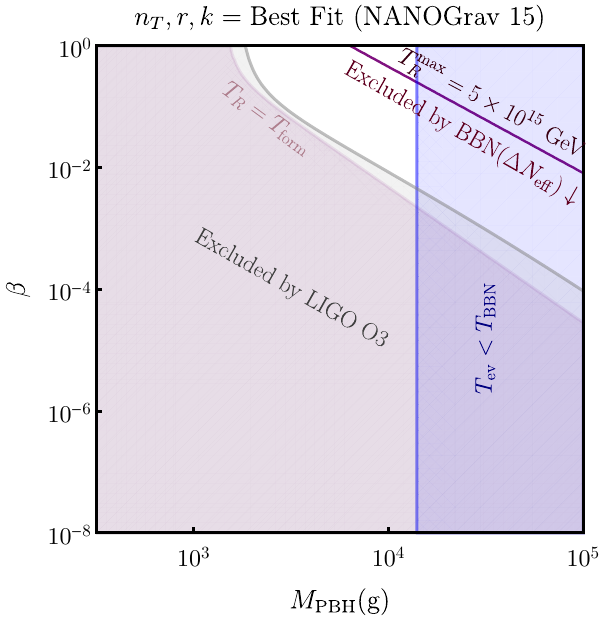}
\includegraphics[width=0.49\textwidth]{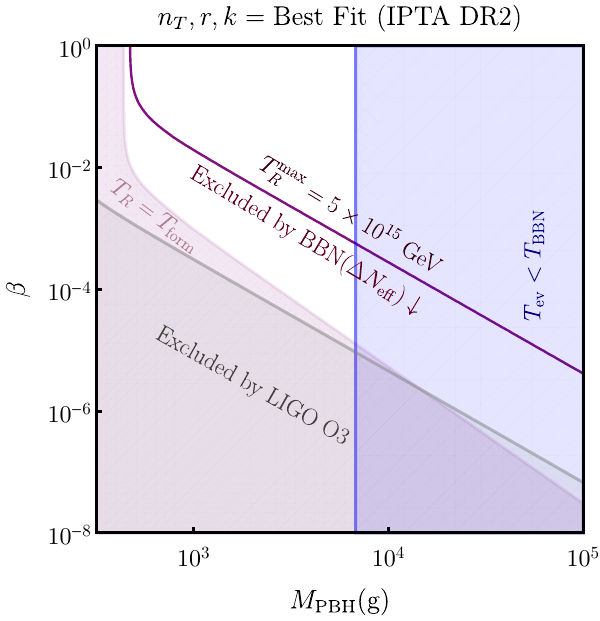}
\caption{Impact of the inflationary reheating temperature $T_{\rm R}$ on the $M_{\rm PBH}$-$\beta$ parameter space, in case of the NANOGrav-15yrs (left panel) and IPTA-DR2 (right panel) best-fits for fixed $q=0.5$ (see Tab.~\ref{t1}).}
\label{fig6}
\end{figure}

Before concluding, several remarks are warranted. Our analysis relies on a power-law parameterisation of the tensor power spectrum as defined in Eq.~\eqref{ps}. Alternative parameterisations~\cite{alt1, alt2} may result in different phenomenological outcomes. Additionally, since we did not assume any specific inflationary model, and as shown in ref. \cite{Agazie:2024kdi}, the minimal Constant Power Law (CPL) model remains sufficient to describe the current NANOGrav signal. While future datasets may allow for a meaningful measurement of nonzero spectral running, at present, its role at PTA frequencies is marginal and does not affect our numerical results regarding PTA constraints on the PBH parameter space. Therefore, we have neglected the running of the spectral index~\cite{Kuroyanagi:2011iw}. For completeness, we provide a discussion of its effects in Appendix \ref{apc}, where we explicitly show how it modifies the SGWB spectrum over a wide frequency range, but this has no impact on our main findings. Our results also show sensitivity to the reheating temperature $T_R$. In Figs.~\ref{fig3} and~\ref{fig5}, we present results for $T_R=T_{\rm form}$. However, we observe that as $T_R$ increases, the allowed parameter regions shrink (see Fig.~\ref{fig6}). Although the predicted GW spectra have a high signal-to-noise ratio (SNR), a thorough spectral shape reconstruction, such as a Fisher analysis~\cite{fish1, fish2}, is required to assess the detectability of subtle spectral features, like $f_{\rm mb}^{\rm kink}$, with confidence. Lastly, it is important to note that PBHs may produce additional GW sources~\cite{Anantua:2008am, Dolgov:2011cq, Dong:2015yjs, Domenech:2020ssp, Domenech:2021wkk, Domenech:2021ztg}; however, a comprehensive study of these contributions lies beyond the scope of this article.

\section{Conclusions \label{sec:concl}}

While ultralight primordial black holes (PBHs) and primordial gravitational waves (GWs) may arise from different cosmic processes, they share a significant connection in the context of recent Pulsar Timing Array (PTA) data, which could provide powerful insights and predictions. Here, we focus on Blue-tilted Gravitational Waves (BGWs), a proposed cosmological candidate for the GWs detected by PTA. However, under the standard cosmological model-- characterized by inflationary reheating, followed by radiation and cold dark matter domination-- BGWs are restricted in their frequency range by the Big Bang Nucleosynthesis (BBN) constraint on the GW amplitude. As a result, detecting BGWs at interferometer scales seems unlikely.

Introducing a phase of early matter domination (EMD) dilutes BGWs at higher frequencies, allowing for consistency with both the BBN and LIGO-O3 constraints on stochastic GWs. This mechanism enables BGWs to align with PTA data while producing a distinctive GW signal observable across a broader frequency spectrum. We propose that ultralight PBHs could provide the necessary EMD phase. Under this framework, PTA results place constraints on the parameter space of the modified Hawking radiation scenario, particularly the ``memory burden" effect associated with ultralight PBHs, offering testable predictions for high-frequency GW detectors.

Notably, we show that, due to the memory burden effect, PBHs as light as $10^2–10^3$ g can leave detectable imprints on BGWs at higher frequencies while still being compatible with PTA data (see Figs.~\ref{fig3} and~\ref{fig5}). Additionally, we observe that the PTA fit is relatively insensitive to the parameter $q$ (see Figs.~\ref{fig2} and~\ref{fig4}), representing the fractional mass retained in PBHs before entering the quantum evaporation phase. However, for smaller $q$ values ($q< 0.5$), memory-burdened PBHs may produce a unique kink-like feature in the GW spectrum (see Fig.~\ref{fig5}).

\section*{Acknowledgements}

The work of both PA and SD is supported by the National Natural Science Foundation of China (NNSFC) under Key Projects grant No. 12335005 and the Research Fund For International Scientists II, grant No. 12150610460.
PA is also supported by the supporting fund for foreign experts grant wgxz2022021.
MC, RS and NS acknowledge the support by the research project TAsP (Theoretical Astroparticle Physics) funded by the Istituto Nazionale di Fisica Nucleare (INFN).
The work of NS is further supported by the research grant number 2022E2J4RK ``PANTHEON: Perspectives in Astroparticle and Neutrino THEory with Old and New messengers'' under the program PRIN 2022 funded by the Italian Ministero dell’Università e della Ricerca (MUR).

\appendix
\section{The memory-burdened PBH evolution sensitivity to $q$}\label{apa}
\begin{figure}
\centering
\includegraphics[width=0.6\textwidth]{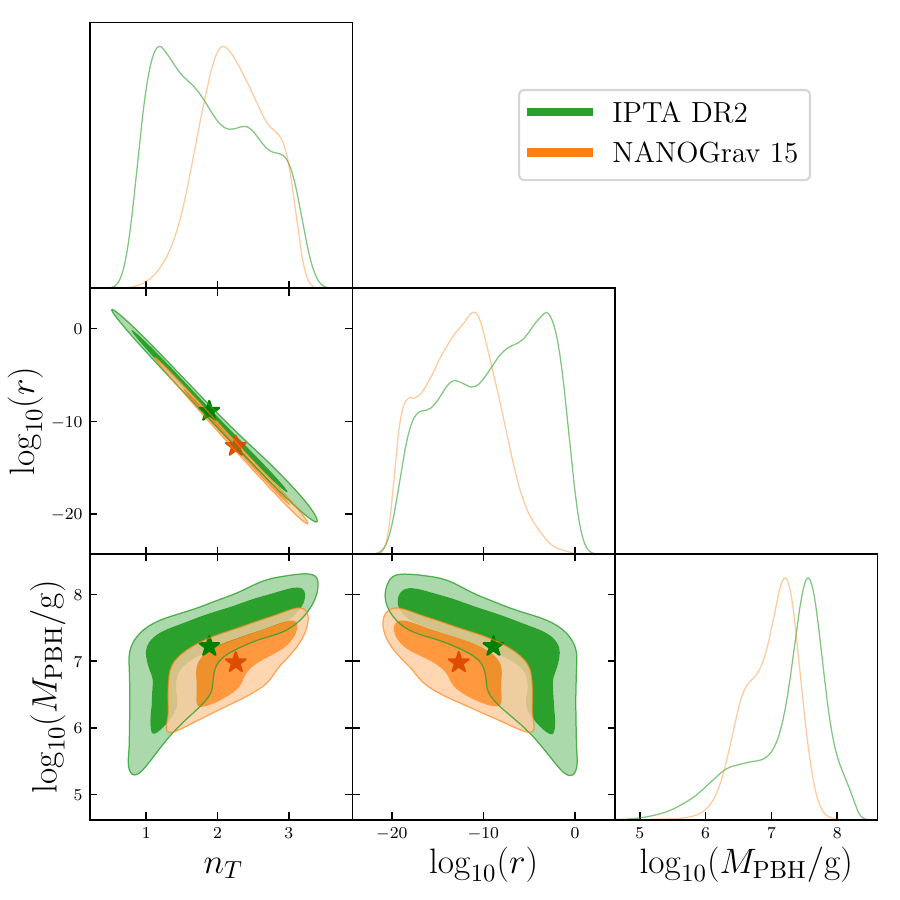}
\caption{Marginalized 2D posterior distributions of the standard PBH and inflationary parameters considering uniform priors for the three parameters. The green (orange) color corresponds to IPTA-DR2 (NANOGrav-15yrs) data, while the different shadings refer to $1\sigma$ and $2\sigma$ regions.}\label{figstd}
\end{figure}

As previously discussed, the memory burden commences when the PBH mass evaporates to $M_{\rm mb} = q M_{\rm PBH}$, with approximately $(1-q)M_{\rm PBH}$ of its energy transforming into radiation. Consequently, the ratio of PBH and radiation energy densities at $T = T_{\rm mb}$ can be expressed as
\begin{eqnarray}
\frac{\rho_{\rm PBH}(T_{\rm mb})}{\rho_{\rm rad}(T_{\rm mb})} &=&\left.\frac{\text{Remanent PBH energy density}}{\text{Radiation energy density} + \text{PBH producing radiation}}\right|_{\text{at }T=T_{\rm mb}}  \nonumber\\
   & = & \frac{\rho_{\rm PBH}(T_{\rm mb})}{\rho_{\rm rad}(T_{\rm mb})+\rho_{\text{PBH}\rightarrow \text{rad}}(T_{\rm mb})} \\
   & = & \frac{q}{(a_{\rm dom}/a_{\rm mb})+(1-q)}\,. \nonumber
\end{eqnarray}
Since we consider the memory burden effect to start significantly after the PBH domination, i.e. $a_{\rm dom}\ll a_{\rm mb}$, there will be a brief period of intermediate radiation domination after the end of the semiclassical Hawking evaporation if
\begin{equation}\label{intrd}
\frac{\rho_{\rm PBH}(T_{\rm mb})}{\rho_{\rm rad}(T_{\rm mb})}< 1 \quad \implies\quad q< 0.5 \footnote{\text{Since this derivation assumes instantaneous energy transfer from PBH to radiation, in practice, $q$ should be} significantly
less than $0.5$ to achieve an intermediate radiation domination.} \,. 
\end{equation}
One can define two periods of entropy dilution due to these two matter-dominated epochs, one due to the evaporation at $T_{\rm mb}$,
\begin{equation}
\Delta_{\rm EMD1}^{\rm sc}=\frac{T_{\rm dom}}{T_{\rm mb}}\,,
\end{equation}
and the other
\begin{equation}
\Delta_{\rm EMD2}^{\rm mb}=\frac{T_{\rm dom2}}{T_{\rm ev}}\,,
\end{equation}
due to the second EMD caused by the memory burden effect, where $T_{\rm dom2}\simeq \left(\frac{q}{1-q}\right) T_{\rm mb}$, taking $q< 0.5$.
In this case, the transfer functions are given by
\begin{equation}
T_T^2(\tau_0,k)=F(k)T_1^2(\zeta_{\rm eq})T_2^2(\zeta_{\text{ev}})T_3^2(\zeta_{\text{\rm dom2}})T_2^2(\zeta_{\rm mb})T_3^2(\zeta_{\rm dom})T_2^2(\zeta_{R}) \,,
\label{transfer}
\end{equation}
whereas the modes $k_i$'s are equal to
\begin{equation}
k_{\text{ev}} = 1.7\times 10^{14}\left(\frac{g_{*s}(T_\text{ev})}{106.75}\right)^{1/6}\left(\frac{T_\text{ev}}{10^7 \rm GeV}\right){\rm Mpc^{-1}}\,,\label{kev} 
\end{equation}
\begin{equation}
k_{\text{dom2}} = 1.7\times 10^{14}(\Delta_{\rm EMD2}^{\rm mb})^{-1/3}\left(\frac{g_{*s}(T_\text{dom2})}{106.75}\right)^{1/6}\left(\frac{T_\text{dom2}}{10^7 \rm GeV}\right){\rm Mpc^{-1}}\,,\label{kdom2} 
\end{equation}
\begin{equation}
k_{\text{mb}} = 1.7\times 10^{14}(\Delta_{\rm EMD2}^{\rm mb})^{-1/3}\left(\frac{g_{*s}(T_\text{mb})}{106.75}\right)^{1/6}\left(\frac{T_\text{mb}}{10^7 \rm GeV}\right){\rm Mpc^{-1}}\,,\label{kmb} 
\end{equation}
\begin{equation}
k_{\text{dom1}} =1.7\times 10^{14}(\Delta_{\rm EMD1}^{\rm sc})^{-1/3} \left(\frac{g_{*s}(T_\text{dom1})}{106.75}\right)^{1/6}\left(\frac{T_\text{dom1}}{10^7 \rm GeV}\right){\rm Mpc^{-1}}\,,\label{kdom}
\end{equation}
\begin{equation}
k_{\rm R}=1.7\times 10^{14}(\Delta_{\rm EMD1}^{\rm sc})^{-1/3}\left(\frac{g_{*s}(T_{\rm R})}{106.75}\right)^{1/6}\left(\frac{T_{\rm R}}{10^7 \rm GeV}\right){\rm Mpc^{-1}}\,.\label{krh}
\end{equation}
In this case, the PBHs initially dominate at $T_{\rm dom1}$, then evaporate at $T_{\rm mb}$ until the threshold mass $M_{\rm mb}$ is reached. We name the region $T_{\rm dom1}>T>T_{\rm mb}$ as EMD1. The intermediate radiation domination occurs for $T_{\rm dom2}<T<T_{\rm mb}$, where $T_{\rm dom2}$ is the onset temperature of the second phase of matter domination (EMD2) which lasts until the memory-burdened PBHs completely evaporate at $T_{\rm ev}$.

\section{Fitting PTA data with semiclassical PBHs}\label{apb}

In Fig.~\ref{figstd}, we show the posteriors of inflationary and PBH parameters for semiclassical PBHs (without the memory burden effect). It is clear that, although BGWs diluted by standard PBH domination fit the PTA data, only ``heavier'' PBHs are allowed, which is in sharp contrast to the memory-burdened case.

\section{Impact of finite running of the tensor spectral index}\label{apc}
To maintain a model-independent approach in our analysis, we examine the impact of a finite running of the tensor spectral index, $\alpha_T=d n_T/ d \ln k$, on the SGWB spectrum. In this scenario, the inflationary tensor power spectrum, $\mathcal{P}_T(k)$, is modified by an additional multiplicative factor given by
\begin{equation}
    \mathcal{P}_T(k) \propto \left(\frac{k}{k_*} \right) ^{\alpha_T \ln(f/f_*)}
\end{equation}
where $f_*$ denotes the frequency at the pivot scale. Notably, for higher frequencies, this correction can significantly influence the SGWB spectrum. Depending on the sign and magnitude of $\alpha_T$, the running can either enhance or suppress the gravitational wave signal at observationally relevant frequencies, introducing an additional source of uncertainty during data interpretation. For a detailed discussion on the effects of running in realistic inflationary models, we refer the readers to ref. \cite{Koh:2018qcy}.\\
\begin{figure}
\centering
\includegraphics[width=0.49\textwidth]{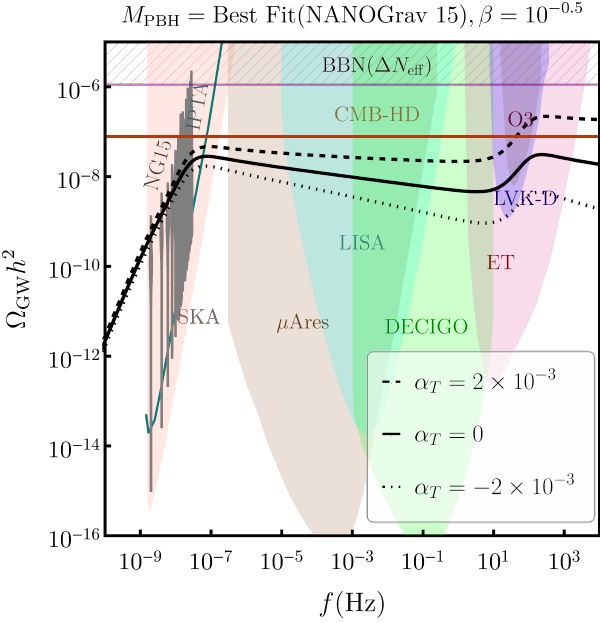}
\includegraphics[width=0.49\textwidth]{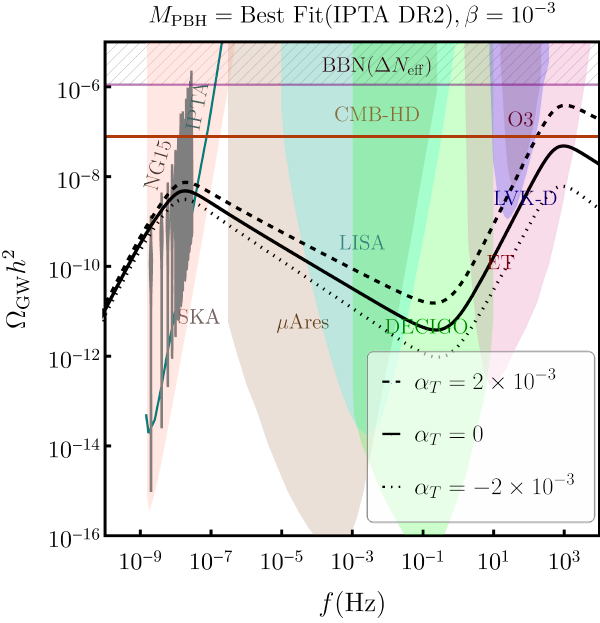}
\caption{Impact of the finite running of the tensor spectral index, $\alpha_T=$ $2\times 10^{-3}$(dashed), $0$ (solid), and $-2\times 10^{-3}$ (dotted), on the BGW spectrum, in case of the NANOGrav-15yrs (left panel) and IPTA-DR2 (right panel) best-fits, with a fixed $q=0.5$ (see Tab.~\ref{t1}).}
\label{fig7}
\end{figure}
In Fig. \ref{fig7}, we present the BGW spectra for different values of $\alpha_T$. It is evident that a nonzero $\alpha_T$ can have a substantial impact on the SGWB spectrum over a wide range of frequencies, introducing an additional parameter that significantly influences the PBH parameter space. However, a detailed investigation of its effects on the PBH parameter space is beyond the scope of this work, as it would require a model-dependent analysis. 
\bibliography{main.bib}
\end{document}